\documentclass[aps,prl,twocolumn,groupedaddress,amsmat,amssymb,amsfonts,superscriptaddress]{revtex4-2}

\usepackage{graphicx}
\usepackage{amsmath,amsfonts,amsthm} 
\usepackage{blkarray}
\usepackage{physics}
\usepackage{mhchem}
\usepackage{caption}
\usepackage{xcolor}
\usepackage{float}
\usepackage{siunitx}

\begin{document}
\raggedbottom
\setlength{\abovedisplayskip}{5pt}
\setlength{\belowdisplayskip}{5pt}
\setlength{\abovedisplayshortskip}{0pt}
\setlength{\belowdisplayshortskip}{0pt}

\title{Transforming from Kitaev to Disguised Ising Chain: Application to CoNb$_2$O$_6$}
\author{Derek Churchill}
\affiliation{Department of Physics, University of Toronto, Ontario, Canada M5S 1A7}
\author{Hae-Young Kee}
\email[]{hy.kee@utoronto.ca}
\affiliation{Department of Physics, University of Toronto, Ontario, Canada M5S 1A7}
\affiliation{Canadian Institute for Advanced Research, CIFAR Program in Quantum Materials, Toronto, Ontario, Canada, M5G 1M1}
\date{\today}

\begin{abstract}
For many years, CoNb$_2$O$_6$ has served as an exemplar 
of the one-dimensional Ising model. However, recent experimental and theoretical analyses challenge its applicability to this material. Prior to that, a tailored spin model for 3d$^7$ systems such as Co$^{2+}$, known as the $JK\Gamma$ model, has emerged, featuring Heisenberg $(J)$, Kitaev $(K)$, and Gamma $(\Gamma)$ interactions. While these interactions are permitted by the symmetry of the system, their role in CoNb$_2$O$_6$ remains enigmatic. 
We present a microscopic theory based on spin-orbit entangled J$_{\rm eff}$ = 1/2 states, aimed at elucidating the roles of Kitaev and Gamma interactions in shaping Ising anisotropy. Leveraging strong coupling theory, we identify a dominant ferromagnetic Kitaev interaction.
Furthermore, 
by comparing dynamical structure factors obtained via exact diagonalization with those from inelastic neutron scattering experiments, we find an antiferromagnetic $\Gamma$ interaction, which dictates the Ising axis and explains the mechanism behind moment pinning.
Our theory suggests that CoNb$_2$O$_6$ represents a rare one-dimensional Kitaev chain with significant ferromagnetic Kitaev and antiferromagnetic Gamma interactions.
\end{abstract}
\maketitle

{\it Introduction -}
The one-dimensional (1D) transverse field Ising model is one of the simplest models exhibiting a quantum phase transition and quantum critical point. The realization of such materials has posed a formidable challenge, with only a limited number of solid-state materials demonstrating a 1D Ising quantum critical point under a magnetic field. Among the extensively studied 1D Ising systems, CoNb$_2$O$_6$ stands out as one of the most investigated examples \cite{Maartense1977SSC,Scharf1979JMMM,Hanawa1994JPS,Heid1995JMMM,Cabrera2014PRB,Coldea2010Sci,Rutkevich2010JSM,Kjall2011PRB,Nandi2019IOP,Fava2020PNAS,Xu2022PRX,Ringler2022PRB,Park2023CGD,Woodland2023PRB2}.

CoNb$_2$O$_6$ exhibits the anticipated quantum critical point under an applied magnetic field of approximately 5T, as verified through inelastic neutron scattering (INS)\cite{Coldea2010Sci}, specific heat\cite{Liang2015NatCom}, THz spectroscopy\cite{Morris2014PRL,Amelin2020PRB,Morris2021Nat,Amelin2022JPA}, and nuclear magnetic resonance measurements\cite{Kinross2014PRX}.
The predicted E$_8$ symmetry\cite{McCoy1978PRD,Zamolodchikov1989IntJ,Delfino1996NucPhyB,Rutkevich2008JSP}, a noteworthy hallmark of the 1D Ising chain near the quantum critical point, has also been identified using INS measurement\cite{Coldea2010Sci}.
However CoNb$_2$O$_6$ displays features inconsistent with those for a pure 1D Ising model with quantum motion of domain walls present even in the absence of an applied transverse field\cite{Coldea2010Sci}. This was recently attributed to a symmetry-allowed staggered spin exchange term\cite{Fava2020PNAS} and a twisted 1D Kitaev model composed of a bond-dependent Ising interaction was suggested to explain features observed in THz spectroscopy measurements in low transverse field\cite{Morris2021Nat}.

Around the same time, the nearest neighbor (n.n.) exchange model of $3d^7$ honeycomb materials was developed \cite{Liu2018PRB,Sano2018PRB}
which is composed of Heisenberg ($J$), bond-dependent Kitaev ($K$), and Gamma ($\Gamma$) interactions known as $JK\Gamma$-model \cite{Khaliullin2005PTR,Kitaev2006AOP,JackeliPRL2009,ChaloupkaPRL2010,Rau2014PRL,Rau2016AnnRev,Winter2017IOP,Motome2020IOP,Liu2020PRL,Winter2022JPM,XLiu2023PRB,Rouso2024RoPP}.
Since Co$^{2+}$ exhibits 3d$^7$ configuration in CoNb$_2$O$_6$, the 1D version of the $JK\Gamma$ model allowed by the system's symmetry is expected.
However, the mechanism by which the $JK\Gamma$-model relates to the Ising anisotropy, their strengths, and the specific roles of each exchange interaction remain unresolved. 

Here we present a microscopic theory of the n.n. exchange interactions to elucidate their roles in Ising anisotropy and domain-wall excitation in the ferromagnetic (FM) ordered state. We find that the FM Kitaev interaction is dominant in CoNb$_2$O$_6$. 
The Ising anisotropy and pinning of the moment direction is due to the antiferromagnetic (AFM) $\Gamma$ interaction. Contribution from other small interactions generated by the octahedra distortion and dynamical structure factor (DSF) obtained via exact diagonalization (ED) are also presented.
  
{\it Microscopic Hamiltonian -}
To derive a microscopic theory, we commence with a brief review of the atomic wave function of Co$^{2+}$, which gives rise to J$_{\rm eff}$=1/2 through the interplay of Hund's coupling and spin-orbit coupling (SOC)
\cite{Liu2018PRB,Sano2018PRB,Liu2020PRL}.
A Co$^{2+}$ ion with a 3d$^7$ electron configuration is surrounded by an octahedral cage of oxygen atoms. 
This generates a cubic crystal field that splits the d-orbital manifold into $t_{2g}$ and $e_{g}$ states, separated by the cubic crystal field energy $\Delta_c$. 
Due to a large Hund's coupling $J_H$ $\left(J_H > \Delta_c\right)$, the Co$^{2+}$ ion forms a high-spin $t_{2g}^5 e_g^2$ electron configuration 
and a 12-fold degenerate $L=1,S=3/2$ subspace is further split by SOC 
resulting in a low-energy, pseudospin-1/2 Kramer's doublet \cite{Liu2018PRB,Sano2018PRB}.

\begin{figure}
  \includegraphics[width=0.48\textwidth]{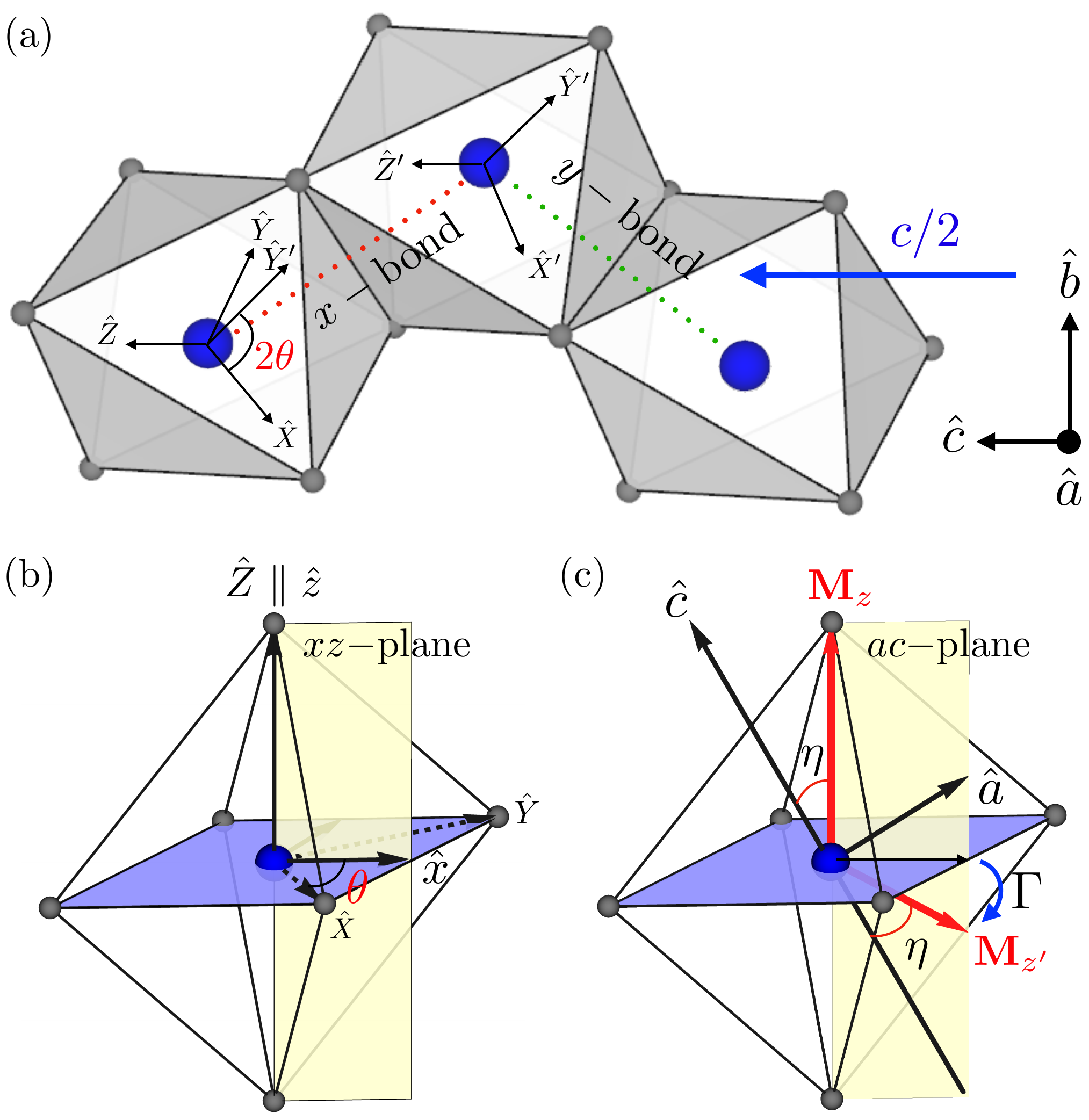}
  \caption{\label{struct} (a)
  Twisted chain consisting of an $x$- and $y$-bond. Each site within the unit cell contains a local cartesian coordiniates, $XYZ$ vs $X'Y'Z'$ (${\hat X}$ and ${\hat Y}'$ are chosen along the anion directions), related by a $c$-glide symmetry along the chain ${\hat c}$ direction.  
  ${\hat Z} \approx {\hat Z}'$ such that the $XY$- and $X'Y'$-plane lie approximately within the same plane, and the angle between ${\hat X}$ and ${\hat Y'}$ axes is defined by $2 \theta$. For an ideal octahedra chain with  $\theta=45^\circ$,  
(b) ${\hat Z}$ lies within the ac-plane, and the global $xyz$-coordinate is shown with respect to the local $XYZ$-coordinate, 
and 
  (c) the red arrows ${\bf M}_z$ and ${\bf M}_{z'}$ in the $ac$-plane represent two magnetic moment directions where $\eta \sim 35^\circ$
  measured from the $\pm {\hat c}$-axis, closer to the experimental findings\cite{Scharf1979JMMM,Heid1995JMMM}.
  }
\end{figure}

The Co$^{2+}$ ions forms a 1D chain 
in the $ac$-plane where the chain is along the crystallographic c-axis as shown in Fig. \ref{struct}(a). They are linked by distorted edge-sharing oxygen octahedra
which define two local cartesian coordinate systems denoted by $XYZ$ and $X'Y'Z'$ respectively, where ${\hat X}$- and ${\hat Y}'$-axis are chosen along their anion directions making the angle of $2\theta$.
Due to the staggered distortion of oxygen cages, C$_2$ symmetries are lost, but there is a glide symmetry with the glide $ac$-plane at $b= 1/4$ as shown in the blue arrow in Fig. \ref{struct}(a).

We first develop a theory for the ideal octahedra cages,
${\hat X}={\hat X}'$, ${\hat Y}={\hat Y}'$ and ${\hat Z}={\hat Z}'$, and the ${\hat Z}$ axis lies within the $ac$-plane.
Using the local $XYZ$ coordinate, the n.n.  generic J$_{\rm eff} =1/2$ exchange model is given by the 
$JK\Gamma$ model\cite{Rau2014PRL,Liu2018PRB,Sano2018PRB,Winter2022JPM,Rouso2024RoPP}:
\begin{gather} 
  H_{ij}^0 = \left[ J \boldsymbol{s}_{i} \cdot \boldsymbol{s}_{j} + 
   K s_{i}^{\gamma} s_{j}^{\gamma} 
   + \Gamma \left(s_{i}^{\alpha} s_{j}^{Z} + s_{i}^Z s_{j}^{\alpha}\right) 
  \right],
\end{gather}
where $\gamma = X(Y)$ and $\alpha = Y(X)$ for the x(y)-bond, 
reflecting the bond-dependent $K$ and $\Gamma$ interactions, and $i j$ refer to the n.n. sites.

It is insightful to rewrite the above model in a global $xyz$ coordinate system. 
In the ideal case, we take ${\hat y} = \frac{1}{\sqrt{2}} [-110]$ 
which is parallel to the ${\hat b}$ axis, and define ${\hat x} = \frac{1}{\sqrt{2}} [110]$ to bisect the ${\hat X}$ and ${\hat Y}$ axis. Then,
${\hat z}$ is fixed to the local oxygen direction ${\hat Z} =  [001]$ as shown in Fig. 1(b).
Transforming the $XYZ$ to the global $xyz$ coordinate, the n.n. Hamiltonian is given by,
\begin{align}\label{nodisteq}
  H^0_{ij}  &=  
  J_K \left(  s_i^x  s_j^x +  s_i^y  s_j^y\right) + J s_i^z s_j^z
+  \frac{\Gamma}{\sqrt{2}} \left(s_i^x s_j^z + s_i^z s_j^x \right) \nonumber\\
   &\hspace{-0.6cm} + (-1)^{i}   \left[ -\frac{K}{2}    \left(s_{i}^x s_j^y + s_{i}^y s_j^x \right)
  +\frac{\Gamma}{\sqrt{2}}   \left(s_i^y s_j^z + s_i^z s_j^y\right)\right],
\end{align}
where $J_{K} = (J + \frac{K}{2})$.

 Before we present the strength of $J$, $K$, and $\Gamma$ interactions, let us discuss the impacts of $K$ and $\Gamma$ in the FM ordered state with $J< 0$. When $K<0$, in the absence of $\Gamma$, the FM moment is pinned via quantum fluctuations along $x \pm y$, which points toward the local ${\hat X}$- or ${\hat Y}$-axis of oxygen direction. As we slightly turn on the AFM $\Gamma$ interaction, the FM moment quickly changes from the ${\hat X}$ (or ${\hat Y}$) to ${\hat x}$ axis, which makes $\eta \sim 55^\circ$ from the $-{\hat c}$-axis. 
 Upon increasing the AFM $\Gamma$ strength, the moment stays in the $ac$-plane but tips closer towards the -${\hat c}$-axis\cite{Chaloupka2016PRB} denoted by the red arrow ${\bf M}_{z'}$.
 Thus a reasonable size of AFM $\Gamma$ interaction is required to achieve the moment direction to be consistent with the experimental finding, $\eta \sim 31^\circ$\cite{Scharf1979JMMM,Heid1995JMMM}.
 On the other hand, if the $K>0$, in the absence of  $\Gamma$, the FM moment is along the local $Z$-axis denoted by the red arrow ${\bf M}_z$ in Fig. 1(c) due to the AFM Kitaev term. 
 This makes the angle of $\eta \sim 35^\circ$ from the $+{\hat c}$ direction. which is already close to the  experimental finding.
 Since $\Gamma$ tips the moment away from the ${\hat z}$-axis\cite{Chaloupka2016PRB}, it implies that $\Gamma \sim 0$ for $K>0$. 
 
 The above analysis for the ideal octahedra cage uncovers that the magnetic moment pinning direction is determined by either AFM Kitaev 
 or a combination of FM Kitaev and the AFM $\Gamma$ interaction. 
 In solid-state materials such as CoNb$_2$O$_6$, the octahedral lattice structure deviates from ideality. The resulting distortion of octahedra induces additional exchange interactions\cite{Liu2020PRL,XLiu2023PRB}, which play a certain role in elucidating the phenomena observed in real materials. Presented below is the complete n.n. Hamiltonian formulation and the methodology employed to estimate these exchange interactions. Our focus lies in identifying the dominant interaction and determining the sign of the Kitaev interaction. This emphasis stems from the fact that the mechanism of the Ising moment pinning hinges on the sign of $K$, which, in turn, informs the role of the Kitaev and Gamma interactions.
 
 {\it Hamiltonian with octahedra distortion} - 
The octahedra distortion modifies the ideal $H^0_{ij}$ and generates other bond dependent interactions such as $K^\prime$, $\Gamma^\prime$, and $\Gamma''$, see the supplementary material (SM) for their definitions in the local $XYZ$ coordinates. Among them $K^\prime$ needs some attention. While $K$ takes the form of $s^x_i s^x_j$ for the x-bond, $K^\prime$ takes the form of $s_i^y s_j^y$ for the x-bond (for y-bond, they are $K s^y_i s^y_j$ and $K^\prime s^x_i s^x_j$). 
Thus when $K=K^\prime$, we lose the bond-dependent Kitaev interaction, and the model with $K=K^\prime$ is nothing but the isotropic XY model. Similarly when $\Gamma = \Gamma^\prime = \Gamma''$, we lose the bond-dependent $\Gamma$ interaction, and it maps to the XXZ model. In general, it requires a fine tuning to make them equal.

In the global coordinates containing the glide-plane as shown in Fig. \ref{struct}(a), $\hat{y}$ is chosen to be parallel to the $\hat{b}$ axis and ${\hat x}$ is chosen to bisect the local ${\hat X}$ and ${\hat Y}'$ axes making an angle $2 \theta$, the n.n.  Hamiltonian $H_{ij}$ including the octahedra distortion in the $xyz$ coordinate system has the following form:
 \begin{gather}
  H_{ij} = 
  J_{xx} s_i^x  s_{j}^x + J_{yy} s_i^y  s_{j}^y 
   + J s_i^z s_{j}^z +  J_{xz} \left(s_i^x s_{j}^z+s_i^z s_{j}^x \right) \notag \\
  \hspace{-0.2cm} \label{fullhameq}+ (-1)^{i}   \biggl[ J_{xy}     \left(s_{i}^x s_{j}^y + s_{i}^y s_{j}^x \right) 
  +J_{yz}   \left(s_i^y s_{i+1}^z + s_i^z s_{i+1}^y\right)\biggr], 
\end{gather}
where, 
\begin{eqnarray}
     J_{xx} &=& \left( J + \frac{ K + K^\prime}{2} +  \Gamma'\right)2 \cos^2(\theta), \label{extinteqs} \nonumber \\  
    J_{yy} &=& \left(J + \frac{ K + K^\prime}{2} -  \Gamma'\right)2\sin^2(\theta), \;\; J_{zz} = J, \nonumber \\
    J_{xz} &= & \left(\Gamma +  \Gamma'' \right) \cos(\theta) \label{jxzeq}, \quad J_{yz} = \left(\Gamma- \Gamma'' \right) \sin(\theta), \nonumber \\
    J_{xy} & =& \frac{( K^\prime-K)}{2} \sin(2\theta).
\end{eqnarray}
Note that the exchange term $J s_i^z s_j^z$ is not modified by the distortion and its strength is determined by the Heisenberg interaction $J$. 
When $\theta = 45^\circ$, i.e., ${\hat x}$ bisects ${\hat X}$ and ${\hat Y} ={\hat Y}'$, the above model is same as the ideal case of $H^0_{ij}$, Eq. \ref{nodisteq}. In CoNb$_2$O$_6$, $\theta \sim 40^\circ$\cite{Heid1995JMMM}.

Let us now estimate the exchange parameters given in Eq. \ref{fullhameq}. 
Due to the complexity of the exchange processes, it is challenging to pin down the numbers below 1 meV precisely using perturbation theory.
Thus we are going to determine the interactions of roughly 1 meV 
using the combination of the density function theory (DFT) and strong coupling expansion. 

{\it Determination of the exchange integrals - }
We estimate the hopping parameters, the crystal field splitting ($\Delta_c$) and charge-transfer gap $(\Delta_{pd})$ using the DFT and maximally localized Wannier functions generated by OpenMX \cite{Ozaki2003PRB,Ozaki2004PRB,Ozaki2005July,Lejaeghere2016Science}. 
The strong coupling expansion requires determining all relevant hopping paths. 
For $d^7$, there are one hole in $t_{2g}$ and two holes in the $e_g$ orbitals, resulting in three different types of exchange paths, $t_{2g}-t_{2g}$, $e_g-t_{2g}$. and $e_g-e_{g}$ processes, denoted as A, B, and C, respectively\cite{Liu2018PRB,XLiu2023PRB}.  Each process includes three different exchanges called intersite-$U$, charge transfer, and cyclic exchanges. Among them 
the C process does not contribute to the Kitaev interaction, since  the $e_g$-$e_g$ exchanges do not change the angular momentum. 

\begin{figure}
    \includegraphics[width=0.9\columnwidth]{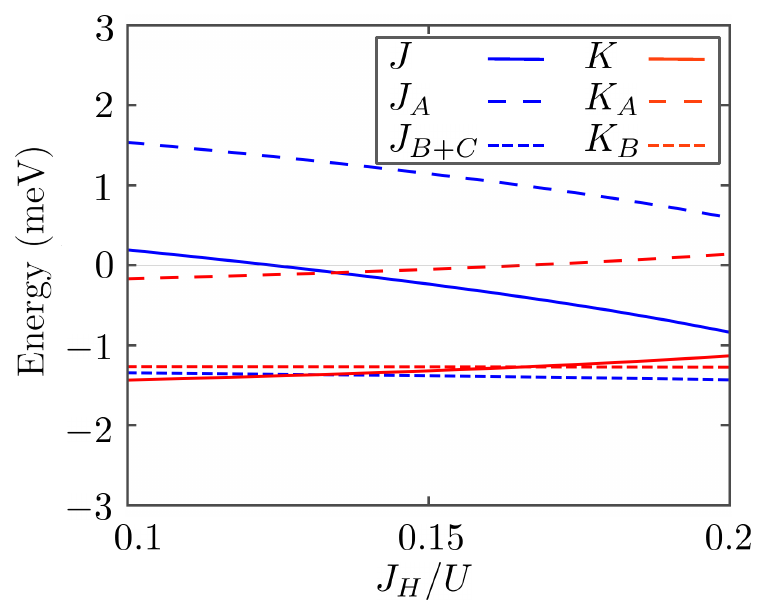}
    \caption{\label{extints} The Heisenberg (blue) and Kitaev (red) interactions vs. $J_H/U$ for $\Delta_{pd}/U=0.365$, $\Delta_c/U =0.09$, and $U =10$ eV. The contributions from $t_{2g}-t_{2g}$ (A), $t_{2g}-e_{g}$ and $e_{g}-e_{g}$ (B+C), and sum of $A+B+C$ processes are represented by the dashed, dotted, and solid lines, respectively. 
    See the main text for details.
    }
\end{figure}

After determining all the relevant hopping parameters using the DFT (see the SM for the tight binding parameters) and then employing the strong coupling expansion, we find the Kitaev interaction is largest for a reasonable size of Hund's coupling.
The dependence of $J$ and $K$ on the Hund's coupling strength in the range of $0.1 < J_H/U < 0.2 $ suitable for cobalt ions\cite{Liu2018PRB} is shown in Fig. \ref{extints}, where A, B, and C processes are plotted using different line styles. The red and blue represent the Kitaev and Heisenberg interactions, respectively. The solid red and blue lines are the total Kitaev and Heisenberg interactions after summing A, B, and C processes, respectively. 
Overall, both the Kitaev and Hisenberg interactions  are FM. 

It is important to note that the Kitaev interaction stays FM and dominant.
This is due to an AFM exchange contribution from $t_{2g}-t_{2g}$ processes cancels with the FM contribution from $e_g-t_{2g}$ and $e_g-e_{g}$ processes, resulting in a small FM $J$. This can be contrasted with 
$3d^7$ honeycomb cobaltates, such as BaCo$_2$(AsO$_4$)$_2$ which has a much larger direct hopping integral leading to the cancellation of Kitaev contribution from
$t_{2g}-t_{2g}$ and $t_{2g}-e_g$ processes yielding
a small Kitaev interaction, but dominant FM Heisenberg interaction.\cite{XLiu2023PRB}
The Kitaev interaction becomes more dominant by moving towards the Mott insulating limit; increasing the charge-transfer gap ($\Delta_{pd}$) while keeping the Hubbard $U$ fixed results in a faster reduction of the FM Heisenberg interaction strength due to a larger AFM contribution from the $t_{2g}-t_{2g}$ processes. 

For $J_H/U = 0.2$,  
we find that $J = -0.8$ meV, $K = -1.1$ meV, and
 $\Gamma$, $K^\prime$, $\Gamma^\prime$, and $\Gamma''$ are small. Since these numbers are sensitive to small changes in the tight binding parameters, 
 we estimate the remaining exchange integrals by computing the DSF from exact-diagonalization (ED), using the open-source numerical package QuSpin \cite{Weinberg2019SPP}, and fit it with the INS data\cite{Woodland2023PRB2}. The details about the DSF computation and systematic fitting procedure to the INS data are explained in the SM. 

A summary of the n.n. exchange parameters is given in Table I and the mapping to $J_{\alpha, \beta}$ with $\alpha, \beta = x,y,z$ is also listed. Note that the $K < 0$ and $J < 0$, while $\Gamma > 0$.  The significant values of $J_{xy}$ and $J_{xz}$ highlight the importance of $K$ and $\Gamma$, as indicated by their relationships in Eq. (4).
With the parameter set listed in Table 1, we found that the Ising moment aligns along the ${\bf M}_{z'}$ direction 
with $\eta \sim 35^\circ$ due to the sizable AFM $\Gamma$ interaction, which causes the moment direction to deviate from the ${\hat x}$-axis towards the $-{\hat c}$-axis, as illustrated by the blue arrow in Fig. 1(c). 

\begin{center}
 \begin{table}[ht]
  \begin{tabular}{| c | c |} 
      \hline
      Interaction in $(XYZ)$ & meV \\
      \hline
          $J$ & -0.8  \\
          $K$ & -1.1 \\
          $\Gamma$ & 0.56\\
          $K^\prime$ & -0.03  \\
          $\Gamma'$ & -0.57  \\
          $\Gamma''$ & 0.26 \\
      \hline
  \end{tabular}
  $\leftrightarrow$
    \begin{tabular}{| c | c |} 
      \hline
      Interaction in $(xyz)$ & meV \\
      \hline
          $J_{xx}$ & -2.27  \\
          $J_{yy}$ & -0.66 \\
          $J_{zz} = J$ & -0.80 \\
          $J_{xy}$ & 0.53  \\
          $J_{xz}$ & 0.63  \\
          $J_{yz}$ & 0.19 \\
      \hline
  \end{tabular}
  \caption{\label{extinttab} Strengths of spin-1/2 n.n. exchange interactions in local $XYZ$ coordinates, and their transformed values in global $xyz$ coordinates.}
  \end{table}
\end{center}

\begin{figure}
    \includegraphics[width=1\columnwidth]{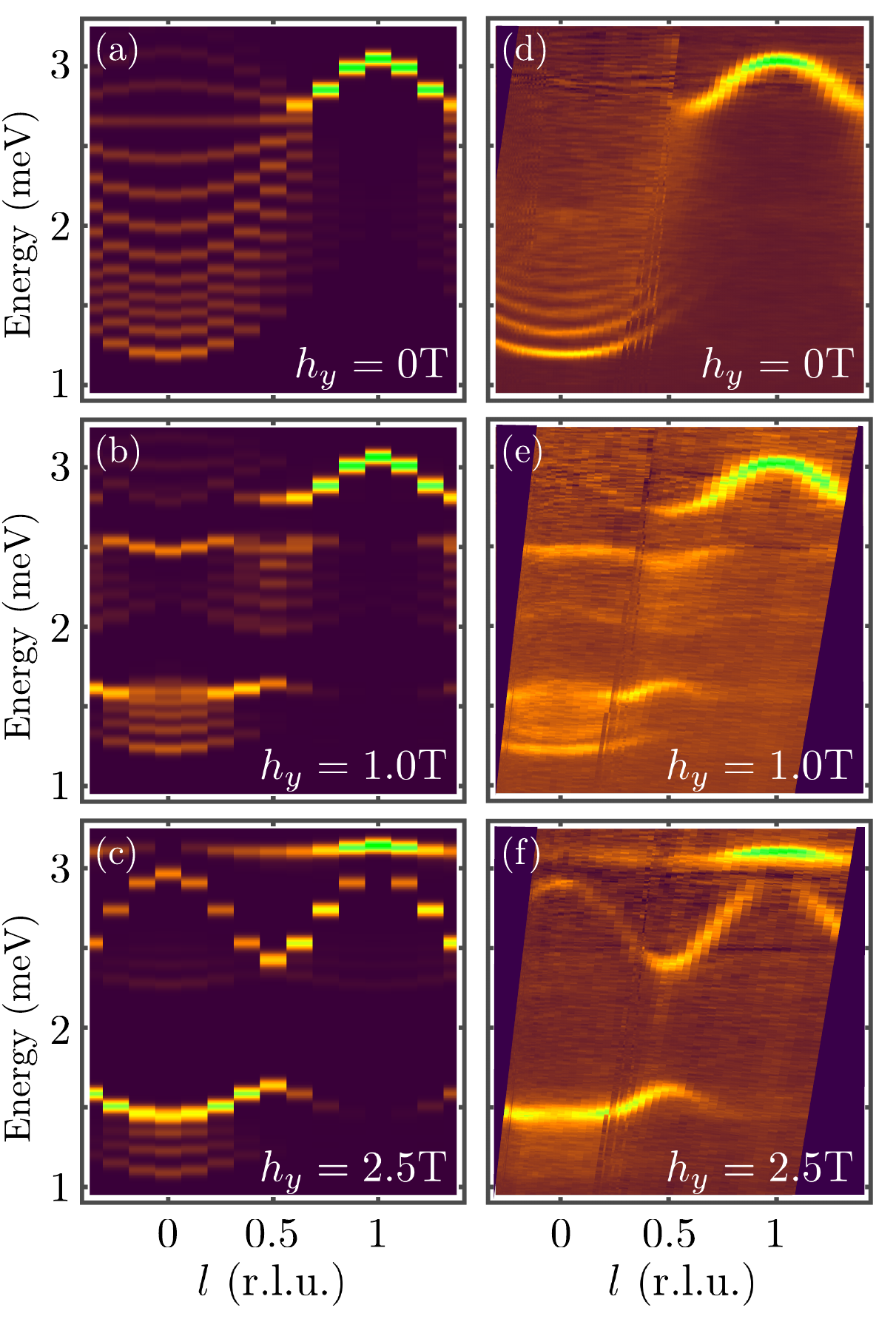}
    \caption{\label{spectralFunction} (a)-(c) $S^{xx}(k,w)$ at various transverse fields obtained by the ED on a 16-site cluster with periodic boundary conditions 
    using the parameters summarized in Table \ref{extinttab} with second n.n. XXZ interactions, a small mean field term, and $g_y = 3.3$ (see the main text for details). (d)-(f) The INS  data adopted from Ref.  \cite{Woodland2023PRB2}.
    }
\end{figure}
\vspace{-0.7cm}

To match the INS data, we include the effects of AFM inter-chain interactions observed in CoNb$_2$O$6$. These interactions confine domain wall pairs, forming bound states. We account for this at the mean field level with a small field term, $\lambda_{\textrm{MF}} \approx 0.04$ meV along the moment direction, and
 also consider second n.n. AFM XXZ interactions as proposed in \cite{Kjall2011PRB}.
We incorporate the transverse field $h_y$ using the g-factor of $g_y = 3.3$.
We then compare the spectral function, with the INS data \cite{Woodland2023PRB2} in Fig. \ref{spectralFunction}, highlighting the close similarity. 

{\it Discussion -} 
The current microscopic theory goes beyond the symmetry-allowed spin model, warranting further discussion. 
Previous analyses, incorporating INS data alongside the symmetry-allowed spin model\cite{Fava2020PNAS,Woodland2023PRB1,Woodland2023PRB2}, empirically choose the Ising axis, thereby impeding the identification of the origin of the Ising anisotropy.
The Ising axis as the moment direction consistent with the experimental results presents
two distinct possibilities: aligning the Ising axis either with ${\bf M}_z$ or ${\bf M}_{z'}$, as depicted by the red arrows in Fig. 1(c).

When the moment aligns with the $z$-axis, directed towards the oxygen atom, it aligns with INS data when the dominant FM $J$ condition holds ($\abs{J} (\equiv |J_{zz}|) > \abs{J_{xx}},\abs{J_{yy}}$). This condition necessitates a significant AFM Kitaev but minimal $\Gamma$ interactions, enforcing $J_{xy} = 0$ and $J_{xz} = 0$ as noted in \cite{Fava2020PNAS,Woodland2023PRB1,Woodland2023PRB2}. Consequently, $K = K'$ and $\Gamma = -\Gamma''$. This, combined with the $J$ term, yields the FM $XXZ$ model with pronounced Ising anisotropy along the local $Z$ axis. However, from a microscopic viewpoint, the equality $K=K'$ is unlikely due to the need for identical contributions from the exchange path of $d_{xz}$-$d_{xy}$ and $d_{yz}$-$d_{xy}$ orbitals for a given bond.
Alternatively, aligning the Ising axis along ${\bf M}_{z'}$ away from the oxygen atom, as found in our theory, offers another approach. Adopting this axis as the empirical Ising axis (denoted by $z'$) allows for the transformation of parameters outlined in Table 1 in the $xyz$ to $x'y'z'$ coordinates while keeping ${\hat y} = {\hat y'}$. The exchange interactions in the $x'y'z'$ coordinate comparing with those in the $xyz$ coordinate are provided in the SM. The resulting n.n. interactions in the $x'y'z'$ coordinate remarkably resemble those reported in \cite{Fava2020PNAS,Woodland2023PRB1,Woodland2023PRB2}, further corroborating our results. 

In summary, through an examination of the exchange interaction derived from the J$_{\rm eff}=1/2$ wavefunction in the local oxygen coordinate, we show that the Ising anisotropy arises from the AFM $\Gamma$ interaction in the presence of the FM Kitaev interaction, which also facilitates domain wall motion. Our theory illustrates that CoNb$_2$O$_6$, once considered an Ising chain exemplar, is a rare bond-dependent Kitaev chain where the Kitaev interaction predominates.
To advance research on two-dimensional Kitaev cobaltates, it is noteworthy to observe the distinction between CoNb$_2$O$_6$ from honeycomb cobaltes such as BaCo$_2$(AsO$_4$)$_2$, where the Kitaev interaction is weakened due to exchange path cancellation\cite{XLiu2023PRB}. This difference is primarily linked to cobalt ion spacing. Our study suggests that increasing the Co-Co distance to minimize direct exchanges is advantageous for enhancing the Kitaev interaction in honeycomb cobaltates, thus facilitating the realization of the Kitaev spin liquid.

{\it Acknowledgments.-} 
We thank P. Armitage, R. Coldea, X. Liu, and L. Woodland for useful discussions.
This work is supported by the NSERC Discovery Grant No. 2022-04601. H.Y.K acknowledges support from the Canada Research Chairs Program. 
Computations were performed on the Niagara supercomputer at
the SciNet HPC Consortium. SciNet is funded by: the
Canada Foundation for Innovation under the auspices of
Compute Canada; the Government of Ontario; Ontario
Research Fund - Research Excellence; and the University
of Toronto.

\bibliography{cno}

\begin{thebibliography}{45}%
\makeatletter
\providecommand \@ifxundefined [1]{%
 \@ifx{#1\undefined}
}%
\providecommand \@ifnum [1]{%
 \ifnum #1\expandafter \@firstoftwo
 \else \expandafter \@secondoftwo
 \fi
}%
\providecommand \@ifx [1]{%
 \ifx #1\expandafter \@firstoftwo
 \else \expandafter \@secondoftwo
 \fi
}%
\providecommand \natexlab [1]{#1}%
\providecommand \enquote  [1]{``#1''}%
\providecommand \bibnamefont  [1]{#1}%
\providecommand \bibfnamefont [1]{#1}%
\providecommand \citenamefont [1]{#1}%
\providecommand \href@noop [0]{\@secondoftwo}%
\providecommand \href [0]{\begingroup \@sanitize@url \@href}%
\providecommand \@href[1]{\@@startlink{#1}\@@href}%
\providecommand \@@href[1]{\endgroup#1\@@endlink}%
\providecommand \@sanitize@url [0]{\catcode `\\12\catcode `\$12\catcode
  `\&12\catcode `\#12\catcode `\^12\catcode `\_12\catcode `\%12\relax}%
\providecommand \@@startlink[1]{}%
\providecommand \@@endlink[0]{}%
\providecommand \url  [0]{\begingroup\@sanitize@url \@url }%
\providecommand \@url [1]{\endgroup\@href {#1}{\urlprefix }}%
\providecommand \urlprefix  [0]{URL }%
\providecommand \Eprint [0]{\href }%
\providecommand \doibase [0]{https://doi.org/}%
\providecommand \selectlanguage [0]{\@gobble}%
\providecommand \bibinfo  [0]{\@secondoftwo}%
\providecommand \bibfield  [0]{\@secondoftwo}%
\providecommand \translation [1]{[#1]}%
\providecommand \BibitemOpen [0]{}%
\providecommand \bibitemStop [0]{}%
\providecommand \bibitemNoStop [0]{.\EOS\space}%
\providecommand \EOS [0]{\spacefactor3000\relax}%
\providecommand \BibitemShut  [1]{\csname bibitem#1\endcsname}%
\let\auto@bib@innerbib\@empty
\bibitem [{\citenamefont {Maartense}\ \emph {et~al.}(1977)\citenamefont
  {Maartense}, \citenamefont {Yaeger},\ and\ \citenamefont
  {Wanklyn}}]{Maartense1977SSC}%
  \BibitemOpen
  \bibfield  {author} {\bibinfo {author} {\bibfnamefont {I.}~\bibnamefont
  {Maartense}}, \bibinfo {author} {\bibfnamefont {I.}~\bibnamefont {Yaeger}},\
  and\ \bibinfo {author} {\bibfnamefont {B.}~\bibnamefont {Wanklyn}},\ }\href
  {https://doi.org/10.1016/0038-1098(77)91485-5} {\bibfield  {journal}
  {\bibinfo  {journal} {Solid State Communications}\ }\textbf {\bibinfo
  {volume} {21}},\ \bibinfo {pages} {93–96} (\bibinfo {year}
  {1977})}\BibitemShut {NoStop}%
\bibitem [{\citenamefont {Scharf}\ \emph {et~al.}(1979)\citenamefont {Scharf},
  \citenamefont {Weitzel}, \citenamefont {Yaeger}, \citenamefont {Maartense},\
  and\ \citenamefont {Wanklyn}}]{Scharf1979JMMM}%
  \BibitemOpen
  \bibfield  {author} {\bibinfo {author} {\bibfnamefont {W.}~\bibnamefont
  {Scharf}}, \bibinfo {author} {\bibfnamefont {H.}~\bibnamefont {Weitzel}},
  \bibinfo {author} {\bibfnamefont {I.}~\bibnamefont {Yaeger}}, \bibinfo
  {author} {\bibfnamefont {I.}~\bibnamefont {Maartense}},\ and\ \bibinfo
  {author} {\bibfnamefont {B.}~\bibnamefont {Wanklyn}},\ }\href
  {https://doi.org/https://doi.org/10.1016/0304-8853(79)90044-1} {\bibfield
  {journal} {\bibinfo  {journal} {Journal of Magnetism and Magnetic Materials}\
  }\textbf {\bibinfo {volume} {13}},\ \bibinfo {pages} {121} (\bibinfo {year}
  {1979})}\BibitemShut {NoStop}%
\bibitem [{\citenamefont {Hanawa}\ \emph {et~al.}(1994)\citenamefont {Hanawa},
  \citenamefont {Shinkawa}, \citenamefont {Ishikawa}, \citenamefont {Miyatani},
  \citenamefont {Saito},\ and\ \citenamefont {Kohn}}]{Hanawa1994JPS}%
  \BibitemOpen
  \bibfield  {author} {\bibinfo {author} {\bibfnamefont {T.}~\bibnamefont
  {Hanawa}}, \bibinfo {author} {\bibfnamefont {K.}~\bibnamefont {Shinkawa}},
  \bibinfo {author} {\bibfnamefont {M.}~\bibnamefont {Ishikawa}}, \bibinfo
  {author} {\bibfnamefont {K.}~\bibnamefont {Miyatani}}, \bibinfo {author}
  {\bibfnamefont {K.}~\bibnamefont {Saito}},\ and\ \bibinfo {author}
  {\bibfnamefont {K.}~\bibnamefont {Kohn}},\ }\href
  {https://doi.org/10.1143/JPSJ.63.2706} {\bibfield  {journal} {\bibinfo
  {journal} {Journal of the Physical Society of Japan}\ }\textbf {\bibinfo
  {volume} {63}},\ \bibinfo {pages} {2706} (\bibinfo {year}
  {1994})}\BibitemShut {NoStop}%
\bibitem [{\citenamefont {Heid}\ \emph {et~al.}(1995)\citenamefont {Heid},
  \citenamefont {Weitzel}, \citenamefont {Burlet}, \citenamefont {Bonnet},
  \citenamefont {Gonschorek}, \citenamefont {Vogt}, \citenamefont {Norwig},\
  and\ \citenamefont {Fuess}}]{Heid1995JMMM}%
  \BibitemOpen
  \bibfield  {author} {\bibinfo {author} {\bibfnamefont {C.}~\bibnamefont
  {Heid}}, \bibinfo {author} {\bibfnamefont {H.}~\bibnamefont {Weitzel}},
  \bibinfo {author} {\bibfnamefont {P.}~\bibnamefont {Burlet}}, \bibinfo
  {author} {\bibfnamefont {M.}~\bibnamefont {Bonnet}}, \bibinfo {author}
  {\bibfnamefont {W.}~\bibnamefont {Gonschorek}}, \bibinfo {author}
  {\bibfnamefont {T.}~\bibnamefont {Vogt}}, \bibinfo {author} {\bibfnamefont
  {J.}~\bibnamefont {Norwig}},\ and\ \bibinfo {author} {\bibfnamefont
  {H.}~\bibnamefont {Fuess}},\ }\href
  {https://doi.org/10.1016/0304-8853(95)00394-0} {\bibfield  {journal}
  {\bibinfo  {journal} {Journal of Magnetism and Magnetic Materials}\ }\textbf
  {\bibinfo {volume} {151}},\ \bibinfo {pages} {123–131} (\bibinfo {year}
  {1995})}\BibitemShut {NoStop}%
\bibitem [{\citenamefont {Cabrera}\ \emph {et~al.}(2014)\citenamefont
  {Cabrera}, \citenamefont {Thompson}, \citenamefont {Coldea}, \citenamefont
  {Prabhakaran}, \citenamefont {Bewley}, \citenamefont {Guidi}, \citenamefont
  {Rodriguez-Rivera},\ and\ \citenamefont {Stock}}]{Cabrera2014PRB}%
  \BibitemOpen
  \bibfield  {author} {\bibinfo {author} {\bibfnamefont {I.}~\bibnamefont
  {Cabrera}}, \bibinfo {author} {\bibfnamefont {J.~D.}\ \bibnamefont
  {Thompson}}, \bibinfo {author} {\bibfnamefont {R.}~\bibnamefont {Coldea}},
  \bibinfo {author} {\bibfnamefont {D.}~\bibnamefont {Prabhakaran}}, \bibinfo
  {author} {\bibfnamefont {R.~I.}\ \bibnamefont {Bewley}}, \bibinfo {author}
  {\bibfnamefont {T.}~\bibnamefont {Guidi}}, \bibinfo {author} {\bibfnamefont
  {J.~A.}\ \bibnamefont {Rodriguez-Rivera}},\ and\ \bibinfo {author}
  {\bibfnamefont {C.}~\bibnamefont {Stock}},\ }\href
  {https://doi.org/10.1103/PhysRevB.90.014418} {\bibfield  {journal} {\bibinfo
  {journal} {Phys. Rev. B}\ }\textbf {\bibinfo {volume} {90}},\ \bibinfo
  {pages} {014418} (\bibinfo {year} {2014})}\BibitemShut {NoStop}%
\bibitem [{\citenamefont {Coldea}\ \emph {et~al.}(2010)\citenamefont {Coldea},
  \citenamefont {Tennant}, \citenamefont {Wheeler}, \citenamefont {Wawrzynska},
  \citenamefont {Prabhakaran}, \citenamefont {Telling}, \citenamefont
  {Habicht}, \citenamefont {Smeibidl},\ and\ \citenamefont
  {Kiefer}}]{Coldea2010Sci}%
  \BibitemOpen
  \bibfield  {author} {\bibinfo {author} {\bibfnamefont {R.}~\bibnamefont
  {Coldea}}, \bibinfo {author} {\bibfnamefont {D.~A.}\ \bibnamefont {Tennant}},
  \bibinfo {author} {\bibfnamefont {E.~M.}\ \bibnamefont {Wheeler}}, \bibinfo
  {author} {\bibfnamefont {E.}~\bibnamefont {Wawrzynska}}, \bibinfo {author}
  {\bibfnamefont {D.}~\bibnamefont {Prabhakaran}}, \bibinfo {author}
  {\bibfnamefont {M.}~\bibnamefont {Telling}}, \bibinfo {author} {\bibfnamefont
  {K.}~\bibnamefont {Habicht}}, \bibinfo {author} {\bibfnamefont
  {P.}~\bibnamefont {Smeibidl}},\ and\ \bibinfo {author} {\bibfnamefont
  {K.}~\bibnamefont {Kiefer}},\ }\href
  {https://doi.org/10.1126/science.1180085} {\bibfield  {journal} {\bibinfo
  {journal} {Science}\ }\textbf {\bibinfo {volume} {327}},\ \bibinfo {pages}
  {177} (\bibinfo {year} {2010})}\BibitemShut {NoStop}%
\bibitem [{\citenamefont {Rutkevich}(2010)}]{Rutkevich2010JSM}%
  \BibitemOpen
  \bibfield  {author} {\bibinfo {author} {\bibfnamefont {S.~B.}\ \bibnamefont
  {Rutkevich}},\ }\href {https://doi.org/10.1088/1742-5468/2010/07/p07015}
  {\bibfield  {journal} {\bibinfo  {journal} {Journal of Statistical Mechanics:
  Theory and Experiment}\ }\textbf {\bibinfo {volume} {2010}},\ \bibinfo
  {pages} {P07015} (\bibinfo {year} {2010})}\BibitemShut {NoStop}%
\bibitem [{\citenamefont {Kjall}\ \emph {et~al.}(2011)\citenamefont {Kjall},
  \citenamefont {Pollmann},\ and\ \citenamefont {Moore}}]{Kjall2011PRB}%
  \BibitemOpen
  \bibfield  {author} {\bibinfo {author} {\bibfnamefont {J.~A.}\ \bibnamefont
  {Kjall}}, \bibinfo {author} {\bibfnamefont {F.}~\bibnamefont {Pollmann}},\
  and\ \bibinfo {author} {\bibfnamefont {J.~E.}\ \bibnamefont {Moore}},\ }\href
  {https://doi.org/10.1103/PhysRevB.83.020407} {\bibfield  {journal} {\bibinfo
  {journal} {Phys. Rev. B}\ }\textbf {\bibinfo {volume} {83}},\ \bibinfo
  {pages} {020407} (\bibinfo {year} {2011})}\BibitemShut {NoStop}%
\bibitem [{\citenamefont {Nandi}\ \emph {et~al.}(2019)\citenamefont {Nandi},
  \citenamefont {Prabhakaran},\ and\ \citenamefont {Mandal}}]{Nandi2019IOP}%
  \BibitemOpen
  \bibfield  {author} {\bibinfo {author} {\bibfnamefont {M.}~\bibnamefont
  {Nandi}}, \bibinfo {author} {\bibfnamefont {D.}~\bibnamefont {Prabhakaran}},\
  and\ \bibinfo {author} {\bibfnamefont {P.}~\bibnamefont {Mandal}},\ }\href
  {https://doi.org/10.1088/1361-648X/ab0539} {\bibfield  {journal} {\bibinfo
  {journal} {Journal of Physics: Condensed Matter}\ }\textbf {\bibinfo {volume}
  {31}},\ \bibinfo {pages} {195802} (\bibinfo {year} {2019})}\BibitemShut
  {NoStop}%
\bibitem [{\citenamefont {Fava}\ \emph {et~al.}(2020)\citenamefont {Fava},
  \citenamefont {Coldea},\ and\ \citenamefont {Parameswaran}}]{Fava2020PNAS}%
  \BibitemOpen
  \bibfield  {author} {\bibinfo {author} {\bibfnamefont {M.}~\bibnamefont
  {Fava}}, \bibinfo {author} {\bibfnamefont {R.}~\bibnamefont {Coldea}},\ and\
  \bibinfo {author} {\bibfnamefont {S.~A.}\ \bibnamefont {Parameswaran}},\
  }\href {https://doi.org/10.1073/pnas.2007986117} {\bibfield  {journal}
  {\bibinfo  {journal} {Proceedings of the National Academy of Sciences}\
  }\textbf {\bibinfo {volume} {117}},\ \bibinfo {pages} {25219–25224}
  (\bibinfo {year} {2020})}\BibitemShut {NoStop}%
\bibitem [{\citenamefont {Xu}\ \emph {et~al.}(2022)\citenamefont {Xu},
  \citenamefont {Wang}, \citenamefont {Huang}, \citenamefont {Ni},
  \citenamefont {Zhao}, \citenamefont {Dai}, \citenamefont {Pan}, \citenamefont
  {Hong}, \citenamefont {Chauhan}, \citenamefont {Koohpayeh}, \citenamefont
  {Armitage},\ and\ \citenamefont {Li}}]{Xu2022PRX}%
  \BibitemOpen
  \bibfield  {author} {\bibinfo {author} {\bibfnamefont {Y.}~\bibnamefont
  {Xu}}, \bibinfo {author} {\bibfnamefont {L.~S.}\ \bibnamefont {Wang}},
  \bibinfo {author} {\bibfnamefont {Y.~Y.}\ \bibnamefont {Huang}}, \bibinfo
  {author} {\bibfnamefont {J.~M.}\ \bibnamefont {Ni}}, \bibinfo {author}
  {\bibfnamefont {C.~C.}\ \bibnamefont {Zhao}}, \bibinfo {author}
  {\bibfnamefont {Y.~F.}\ \bibnamefont {Dai}}, \bibinfo {author} {\bibfnamefont
  {B.~Y.}\ \bibnamefont {Pan}}, \bibinfo {author} {\bibfnamefont {X.~C.}\
  \bibnamefont {Hong}}, \bibinfo {author} {\bibfnamefont {P.}~\bibnamefont
  {Chauhan}}, \bibinfo {author} {\bibfnamefont {S.~M.}\ \bibnamefont
  {Koohpayeh}}, \bibinfo {author} {\bibfnamefont {N.~P.}\ \bibnamefont
  {Armitage}},\ and\ \bibinfo {author} {\bibfnamefont {S.~Y.}\ \bibnamefont
  {Li}},\ }\href {https://doi.org/10.1103/PhysRevX.12.021020} {\bibfield
  {journal} {\bibinfo  {journal} {Phys. Rev. X}\ }\textbf {\bibinfo {volume}
  {12}},\ \bibinfo {pages} {021020} (\bibinfo {year} {2022})}\BibitemShut
  {NoStop}%
\bibitem [{\citenamefont {Ringler}\ \emph {et~al.}(2022)\citenamefont
  {Ringler}, \citenamefont {Kolesnikov},\ and\ \citenamefont
  {Ross}}]{Ringler2022PRB}%
  \BibitemOpen
  \bibfield  {author} {\bibinfo {author} {\bibfnamefont {J.~A.}\ \bibnamefont
  {Ringler}}, \bibinfo {author} {\bibfnamefont {A.~I.}\ \bibnamefont
  {Kolesnikov}},\ and\ \bibinfo {author} {\bibfnamefont {K.~A.}\ \bibnamefont
  {Ross}},\ }\href {https://doi.org/10.1103/PhysRevB.105.224421} {\bibfield
  {journal} {\bibinfo  {journal} {Phys. Rev. B}\ }\textbf {\bibinfo {volume}
  {105}},\ \bibinfo {pages} {224421} (\bibinfo {year} {2022})}\BibitemShut
  {NoStop}%
\bibitem [{\citenamefont {Park}\ \emph {et~al.}()\citenamefont {Park},
  \citenamefont {Lee}, \citenamefont {Bae}, \citenamefont {Kim},\ and\
  \citenamefont {Yoon}}]{Park2023CGD}%
  \BibitemOpen
  \bibfield  {author} {\bibinfo {author} {\bibfnamefont {J.}~\bibnamefont
  {Park}}, \bibinfo {author} {\bibfnamefont {S.~H.}\ \bibnamefont {Lee}},
  \bibinfo {author} {\bibfnamefont {S.}~\bibnamefont {Bae}}, \bibinfo {author}
  {\bibfnamefont {M.~H.}\ \bibnamefont {Kim}},\ and\ \bibinfo {author}
  {\bibfnamefont {S.}~\bibnamefont {Yoon}},\ }\href@noop {} {\ }\BibitemShut
  {NoStop}%
\bibitem [{\citenamefont {Woodland}\ \emph
  {et~al.}(2023{\natexlab{a}})\citenamefont {Woodland}, \citenamefont {Lovas},
  \citenamefont {Telling}, \citenamefont {Prabhakaran}, \citenamefont
  {Balents},\ and\ \citenamefont {Coldea}}]{Woodland2023PRB2}%
  \BibitemOpen
  \bibfield  {author} {\bibinfo {author} {\bibfnamefont {L.}~\bibnamefont
  {Woodland}}, \bibinfo {author} {\bibfnamefont {I.}~\bibnamefont {Lovas}},
  \bibinfo {author} {\bibfnamefont {M.}~\bibnamefont {Telling}}, \bibinfo
  {author} {\bibfnamefont {D.}~\bibnamefont {Prabhakaran}}, \bibinfo {author}
  {\bibfnamefont {L.}~\bibnamefont {Balents}},\ and\ \bibinfo {author}
  {\bibfnamefont {R.}~\bibnamefont {Coldea}},\ }\href
  {https://doi.org/10.1103/PhysRevB.108.184417} {\bibfield  {journal} {\bibinfo
   {journal} {Phys. Rev. B}\ }\textbf {\bibinfo {volume} {108}},\ \bibinfo
  {pages} {184417} (\bibinfo {year} {2023}{\natexlab{a}})}\BibitemShut
  {NoStop}%
\bibitem [{\citenamefont {Liang}\ \emph {et~al.}(2015)\citenamefont {Liang},
  \citenamefont {Koohpayeh}, \citenamefont {Krizan}, \citenamefont {McQueen},
  \citenamefont {Cava},\ and\ \citenamefont {Ong}}]{Liang2015NatCom}%
  \BibitemOpen
  \bibfield  {author} {\bibinfo {author} {\bibfnamefont {T.}~\bibnamefont
  {Liang}}, \bibinfo {author} {\bibfnamefont {S.~M.}\ \bibnamefont
  {Koohpayeh}}, \bibinfo {author} {\bibfnamefont {J.~W.}\ \bibnamefont
  {Krizan}}, \bibinfo {author} {\bibfnamefont {T.~M.}\ \bibnamefont {McQueen}},
  \bibinfo {author} {\bibfnamefont {R.~J.}\ \bibnamefont {Cava}},\ and\
  \bibinfo {author} {\bibfnamefont {N.~P.}\ \bibnamefont {Ong}},\ }\href
  {https://doi.org/10.1038/ncomms8611} {\bibfield  {journal} {\bibinfo
  {journal} {Nature Communications}\ }\textbf {\bibinfo {volume} {6}},\
  \bibinfo {pages} {7611} (\bibinfo {year} {2015})}\BibitemShut {NoStop}%
\bibitem [{\citenamefont {Morris}\ \emph {et~al.}(2014)\citenamefont {Morris},
  \citenamefont {Vald\'es~Aguilar}, \citenamefont {Ghosh}, \citenamefont
  {Koohpayeh}, \citenamefont {Krizan}, \citenamefont {Cava}, \citenamefont
  {Tchernyshyov}, \citenamefont {McQueen},\ and\ \citenamefont
  {Armitage}}]{Morris2014PRL}%
  \BibitemOpen
  \bibfield  {author} {\bibinfo {author} {\bibfnamefont {C.~M.}\ \bibnamefont
  {Morris}}, \bibinfo {author} {\bibfnamefont {R.}~\bibnamefont
  {Vald\'es~Aguilar}}, \bibinfo {author} {\bibfnamefont {A.}~\bibnamefont
  {Ghosh}}, \bibinfo {author} {\bibfnamefont {S.~M.}\ \bibnamefont
  {Koohpayeh}}, \bibinfo {author} {\bibfnamefont {J.}~\bibnamefont {Krizan}},
  \bibinfo {author} {\bibfnamefont {R.~J.}\ \bibnamefont {Cava}}, \bibinfo
  {author} {\bibfnamefont {O.}~\bibnamefont {Tchernyshyov}}, \bibinfo {author}
  {\bibfnamefont {T.~M.}\ \bibnamefont {McQueen}},\ and\ \bibinfo {author}
  {\bibfnamefont {N.~P.}\ \bibnamefont {Armitage}},\ }\href
  {https://doi.org/10.1103/PhysRevLett.112.137403} {\bibfield  {journal}
  {\bibinfo  {journal} {Phys. Rev. Lett.}\ }\textbf {\bibinfo {volume} {112}},\
  \bibinfo {pages} {137403} (\bibinfo {year} {2014})}\BibitemShut {NoStop}%
\bibitem [{\citenamefont {Amelin}\ \emph {et~al.}(2020)\citenamefont {Amelin},
  \citenamefont {Engelmayer}, \citenamefont {Viirok}, \citenamefont {Nagel},
  \citenamefont {R\~o\ om}, \citenamefont {Lorenz},\ and\ \citenamefont
  {Wang}}]{Amelin2020PRB}%
  \BibitemOpen
  \bibfield  {author} {\bibinfo {author} {\bibfnamefont {K.}~\bibnamefont
  {Amelin}}, \bibinfo {author} {\bibfnamefont {J.}~\bibnamefont {Engelmayer}},
  \bibinfo {author} {\bibfnamefont {J.}~\bibnamefont {Viirok}}, \bibinfo
  {author} {\bibfnamefont {U.}~\bibnamefont {Nagel}}, \bibinfo {author}
  {\bibfnamefont {T.}~\bibnamefont {R\~o\ om}}, \bibinfo {author}
  {\bibfnamefont {T.}~\bibnamefont {Lorenz}},\ and\ \bibinfo {author}
  {\bibfnamefont {Z.}~\bibnamefont {Wang}},\ }\href
  {https://doi.org/10.1103/PhysRevB.102.104431} {\bibfield  {journal} {\bibinfo
   {journal} {Phys. Rev. B}\ }\textbf {\bibinfo {volume} {102}},\ \bibinfo
  {pages} {104431} (\bibinfo {year} {2020})}\BibitemShut {NoStop}%
\bibitem [{\citenamefont {Morris}\ \emph {et~al.}(2021)\citenamefont {Morris},
  \citenamefont {Desai}, \citenamefont {Viirok}, \citenamefont {H{\"u}vonen},
  \citenamefont {Nagel}, \citenamefont {R{\~o}{\~o}m}, \citenamefont {Krizan},
  \citenamefont {Cava}, \citenamefont {McQueen}, \citenamefont {Koohpayeh},
  \citenamefont {Kaul},\ and\ \citenamefont {Armitage}}]{Morris2021Nat}%
  \BibitemOpen
  \bibfield  {author} {\bibinfo {author} {\bibfnamefont {C.~M.}\ \bibnamefont
  {Morris}}, \bibinfo {author} {\bibfnamefont {N.}~\bibnamefont {Desai}},
  \bibinfo {author} {\bibfnamefont {J.}~\bibnamefont {Viirok}}, \bibinfo
  {author} {\bibfnamefont {D.}~\bibnamefont {H{\"u}vonen}}, \bibinfo {author}
  {\bibfnamefont {U.}~\bibnamefont {Nagel}}, \bibinfo {author} {\bibfnamefont
  {T.}~\bibnamefont {R{\~o}{\~o}m}}, \bibinfo {author} {\bibfnamefont {J.~W.}\
  \bibnamefont {Krizan}}, \bibinfo {author} {\bibfnamefont {R.~J.}\
  \bibnamefont {Cava}}, \bibinfo {author} {\bibfnamefont {T.~M.}\ \bibnamefont
  {McQueen}}, \bibinfo {author} {\bibfnamefont {S.~M.}\ \bibnamefont
  {Koohpayeh}}, \bibinfo {author} {\bibfnamefont {R.~K.}\ \bibnamefont
  {Kaul}},\ and\ \bibinfo {author} {\bibfnamefont {N.~P.}\ \bibnamefont
  {Armitage}},\ }\href {https://doi.org/10.1038/s41567-021-01208-0} {\bibfield
  {journal} {\bibinfo  {journal} {Nature Physics}\ }\textbf {\bibinfo {volume}
  {17}},\ \bibinfo {pages} {832} (\bibinfo {year} {2021})}\BibitemShut
  {NoStop}%
\bibitem [{\citenamefont {Amelin}\ \emph {et~al.}(2022)\citenamefont {Amelin},
  \citenamefont {Viirok}, \citenamefont {Nagel}, \citenamefont {Rõõm},
  \citenamefont {Engelmayer}, \citenamefont {Dey}, \citenamefont
  {Agung~Nugroho}, \citenamefont {Lorenz},\ and\ \citenamefont
  {Wang}}]{Amelin2022JPA}%
  \BibitemOpen
  \bibfield  {author} {\bibinfo {author} {\bibfnamefont {K.}~\bibnamefont
  {Amelin}}, \bibinfo {author} {\bibfnamefont {J.}~\bibnamefont {Viirok}},
  \bibinfo {author} {\bibfnamefont {U.}~\bibnamefont {Nagel}}, \bibinfo
  {author} {\bibfnamefont {T.}~\bibnamefont {Rõõm}}, \bibinfo {author}
  {\bibfnamefont {J.}~\bibnamefont {Engelmayer}}, \bibinfo {author}
  {\bibfnamefont {T.}~\bibnamefont {Dey}}, \bibinfo {author} {\bibfnamefont
  {A.}~\bibnamefont {Agung~Nugroho}}, \bibinfo {author} {\bibfnamefont
  {T.}~\bibnamefont {Lorenz}},\ and\ \bibinfo {author} {\bibfnamefont
  {Z.}~\bibnamefont {Wang}},\ }\href {https://doi.org/10.1088/1751-8121/aca6b8}
  {\bibfield  {journal} {\bibinfo  {journal} {Journal of Physics A:
  Mathematical and Theoretical}\ }\textbf {\bibinfo {volume} {55}},\ \bibinfo
  {pages} {484005} (\bibinfo {year} {2022})}\BibitemShut {NoStop}%
\bibitem [{\citenamefont {Kinross}\ \emph {et~al.}(2014)\citenamefont
  {Kinross}, \citenamefont {Fu}, \citenamefont {Munsie}, \citenamefont
  {Dabkowska}, \citenamefont {Luke}, \citenamefont {Sachdev},\ and\
  \citenamefont {Imai}}]{Kinross2014PRX}%
  \BibitemOpen
  \bibfield  {author} {\bibinfo {author} {\bibfnamefont {A.~W.}\ \bibnamefont
  {Kinross}}, \bibinfo {author} {\bibfnamefont {M.}~\bibnamefont {Fu}},
  \bibinfo {author} {\bibfnamefont {T.~J.}\ \bibnamefont {Munsie}}, \bibinfo
  {author} {\bibfnamefont {H.~A.}\ \bibnamefont {Dabkowska}}, \bibinfo {author}
  {\bibfnamefont {G.~M.}\ \bibnamefont {Luke}}, \bibinfo {author}
  {\bibfnamefont {S.}~\bibnamefont {Sachdev}},\ and\ \bibinfo {author}
  {\bibfnamefont {T.}~\bibnamefont {Imai}},\ }\href
  {https://doi.org/10.1103/PhysRevX.4.031008} {\bibfield  {journal} {\bibinfo
  {journal} {Phys. Rev. X}\ }\textbf {\bibinfo {volume} {4}},\ \bibinfo {pages}
  {031008} (\bibinfo {year} {2014})}\BibitemShut {NoStop}%
\bibitem [{\citenamefont {McCoy}\ and\ \citenamefont
  {Wu}(1978)}]{McCoy1978PRD}%
  \BibitemOpen
  \bibfield  {author} {\bibinfo {author} {\bibfnamefont {B.~M.}\ \bibnamefont
  {McCoy}}\ and\ \bibinfo {author} {\bibfnamefont {T.~T.}\ \bibnamefont {Wu}},\
  }\href {https://doi.org/10.1103/PhysRevD.18.1259} {\bibfield  {journal}
  {\bibinfo  {journal} {Phys. Rev. D}\ }\textbf {\bibinfo {volume} {18}},\
  \bibinfo {pages} {1259} (\bibinfo {year} {1978})}\BibitemShut {NoStop}%
\bibitem [{\citenamefont {Zamolodchikov}(1989)}]{Zamolodchikov1989IntJ}%
  \BibitemOpen
  \bibfield  {author} {\bibinfo {author} {\bibfnamefont {A.~B.}\ \bibnamefont
  {Zamolodchikov}},\ }\href {https://doi.org/10.1142/S0217751X8900176X}
  {\bibfield  {journal} {\bibinfo  {journal} {Int. J. Mod. Phys. A}\ }\textbf
  {\bibinfo {volume} {4}},\ \bibinfo {pages} {4235} (\bibinfo {year}
  {1989})}\BibitemShut {NoStop}%
\bibitem [{\citenamefont {Delfino}\ \emph {et~al.}(1996)\citenamefont
  {Delfino}, \citenamefont {Mussardo},\ and\ \citenamefont
  {Simonetti}}]{Delfino1996NucPhyB}%
  \BibitemOpen
  \bibfield  {author} {\bibinfo {author} {\bibfnamefont {G.}~\bibnamefont
  {Delfino}}, \bibinfo {author} {\bibfnamefont {G.}~\bibnamefont {Mussardo}},\
  and\ \bibinfo {author} {\bibfnamefont {P.}~\bibnamefont {Simonetti}},\ }\href
  {https://doi.org/10.1016/0550-3213(96)00265-9} {\bibfield  {journal}
  {\bibinfo  {journal} {Nucl. Phys. B}\ }\textbf {\bibinfo {volume} {473}},\
  \bibinfo {pages} {469} (\bibinfo {year} {1996})}\BibitemShut {NoStop}%
\bibitem [{\citenamefont {Rutkevich}(2008)}]{Rutkevich2008JSP}%
  \BibitemOpen
  \bibfield  {author} {\bibinfo {author} {\bibfnamefont {S.~B.}\ \bibnamefont
  {Rutkevich}},\ }\href {https://doi.org/10.1007/s10955-008-9495-1} {\bibfield
  {journal} {\bibinfo  {journal} {Journal of Statistical Physics}\ }\textbf
  {\bibinfo {volume} {131}},\ \bibinfo {pages} {917} (\bibinfo {year}
  {2008})}\BibitemShut {NoStop}%
\bibitem [{\citenamefont {Liu}\ and\ \citenamefont
  {Khaliullin}(2018)}]{Liu2018PRB}%
  \BibitemOpen
  \bibfield  {author} {\bibinfo {author} {\bibfnamefont {H.}~\bibnamefont
  {Liu}}\ and\ \bibinfo {author} {\bibfnamefont {G.}~\bibnamefont
  {Khaliullin}},\ }\href {https://doi.org/10.1103/PhysRevB.97.014407}
  {\bibfield  {journal} {\bibinfo  {journal} {Phys. Rev. B}\ }\textbf {\bibinfo
  {volume} {97}},\ \bibinfo {pages} {014407} (\bibinfo {year}
  {2018})}\BibitemShut {NoStop}%
\bibitem [{\citenamefont {Sano}\ \emph {et~al.}(2018)\citenamefont {Sano},
  \citenamefont {Kato},\ and\ \citenamefont {Motome}}]{Sano2018PRB}%
  \BibitemOpen
  \bibfield  {author} {\bibinfo {author} {\bibfnamefont {R.}~\bibnamefont
  {Sano}}, \bibinfo {author} {\bibfnamefont {Y.}~\bibnamefont {Kato}},\ and\
  \bibinfo {author} {\bibfnamefont {Y.}~\bibnamefont {Motome}},\ }\href
  {https://doi.org/10.1103/PhysRevB.97.014408} {\bibfield  {journal} {\bibinfo
  {journal} {Phys. Rev. B}\ }\textbf {\bibinfo {volume} {97}},\ \bibinfo
  {pages} {014408} (\bibinfo {year} {2018})}\BibitemShut {NoStop}%
\bibitem [{\citenamefont {Khaliullin}(2005)}]{Khaliullin2005PTR}%
  \BibitemOpen
  \bibfield  {author} {\bibinfo {author} {\bibfnamefont {G.}~\bibnamefont
  {Khaliullin}},\ }\href {https://doi.org/10.1143/PTPS.160.155} {\bibfield
  {journal} {\bibinfo  {journal} {Progress of Theoretical Physics Supplement}\
  }\textbf {\bibinfo {volume} {160}},\ \bibinfo {pages} {155} (\bibinfo {year}
  {2005})}\BibitemShut {NoStop}%
\bibitem [{\citenamefont {Kitaev}(2006)}]{Kitaev2006AOP}%
  \BibitemOpen
  \bibfield  {author} {\bibinfo {author} {\bibfnamefont {A.}~\bibnamefont
  {Kitaev}},\ }\href
  {https://doi.org/https://doi.org/10.1016/j.aop.2005.10.005} {\bibfield
  {journal} {\bibinfo  {journal} {Annals of Physics}\ }\textbf {\bibinfo
  {volume} {321}},\ \bibinfo {pages} {2} (\bibinfo {year} {2006})},\ \bibinfo
  {note} {january Special Issue}\BibitemShut {NoStop}%
\bibitem [{\citenamefont {Jackeli}\ and\ \citenamefont
  {Khaliullin}(2009)}]{JackeliPRL2009}%
  \BibitemOpen
  \bibfield  {author} {\bibinfo {author} {\bibfnamefont {G.}~\bibnamefont
  {Jackeli}}\ and\ \bibinfo {author} {\bibfnamefont {G.}~\bibnamefont
  {Khaliullin}},\ }\href {https://doi.org/10.1103/PhysRevLett.102.017205}
  {\bibfield  {journal} {\bibinfo  {journal} {Phys. Rev. Lett.}\ }\textbf
  {\bibinfo {volume} {102}},\ \bibinfo {pages} {017205} (\bibinfo {year}
  {2009})}\BibitemShut {NoStop}%
\bibitem [{\citenamefont {Chaloupka}\ \emph {et~al.}(2010)\citenamefont
  {Chaloupka}, \citenamefont {Jackeli},\ and\ \citenamefont
  {Khaliullin}}]{ChaloupkaPRL2010}%
  \BibitemOpen
  \bibfield  {author} {\bibinfo {author} {\bibfnamefont {J.~c.~v.}\
  \bibnamefont {Chaloupka}}, \bibinfo {author} {\bibfnamefont {G.}~\bibnamefont
  {Jackeli}},\ and\ \bibinfo {author} {\bibfnamefont {G.}~\bibnamefont
  {Khaliullin}},\ }\href {https://doi.org/10.1103/PhysRevLett.105.027204}
  {\bibfield  {journal} {\bibinfo  {journal} {Phys. Rev. Lett.}\ }\textbf
  {\bibinfo {volume} {105}},\ \bibinfo {pages} {027204} (\bibinfo {year}
  {2010})}\BibitemShut {NoStop}%
\bibitem [{\citenamefont {Rau}\ \emph {et~al.}(2014)\citenamefont {Rau},
  \citenamefont {Lee},\ and\ \citenamefont {Kee}}]{Rau2014PRL}%
  \BibitemOpen
  \bibfield  {author} {\bibinfo {author} {\bibfnamefont {J.~G.}\ \bibnamefont
  {Rau}}, \bibinfo {author} {\bibfnamefont {E.~K.-H.}\ \bibnamefont {Lee}},\
  and\ \bibinfo {author} {\bibfnamefont {H.-Y.}\ \bibnamefont {Kee}},\ }\href
  {https://doi.org/10.1103/PhysRevLett.112.077204} {\bibfield  {journal}
  {\bibinfo  {journal} {Phys. Rev. Lett.}\ }\textbf {\bibinfo {volume} {112}},\
  \bibinfo {pages} {077204} (\bibinfo {year} {2014})}\BibitemShut {NoStop}%
\bibitem [{\citenamefont {Rau}\ \emph {et~al.}(2016)\citenamefont {Rau},
  \citenamefont {Lee},\ and\ \citenamefont {Kee}}]{Rau2016AnnRev}%
  \BibitemOpen
  \bibfield  {author} {\bibinfo {author} {\bibfnamefont {J.~G.}\ \bibnamefont
  {Rau}}, \bibinfo {author} {\bibfnamefont {E.~K.-H.}\ \bibnamefont {Lee}},\
  and\ \bibinfo {author} {\bibfnamefont {H.-Y.}\ \bibnamefont {Kee}},\ }\href
  {https://doi.org/10.1146/annurev-conmatphys-031115-011319} {\bibfield
  {journal} {\bibinfo  {journal} {Annual Review of Condensed Matter Physics}\
  }\textbf {\bibinfo {volume} {7}},\ \bibinfo {pages} {195} (\bibinfo {year}
  {2016})}\BibitemShut {NoStop}%
\bibitem [{\citenamefont {Winter}\ \emph {et~al.}(2017)\citenamefont {Winter},
  \citenamefont {Tsirlin}, \citenamefont {Daghofer}, \citenamefont {van~den
  Brink}, \citenamefont {Singh}, \citenamefont {Gegenwart},\ and\ \citenamefont
  {Valentí}}]{Winter2017IOP}%
  \BibitemOpen
  \bibfield  {author} {\bibinfo {author} {\bibfnamefont {S.~M.}\ \bibnamefont
  {Winter}}, \bibinfo {author} {\bibfnamefont {A.~A.}\ \bibnamefont {Tsirlin}},
  \bibinfo {author} {\bibfnamefont {M.}~\bibnamefont {Daghofer}}, \bibinfo
  {author} {\bibfnamefont {J.}~\bibnamefont {van~den Brink}}, \bibinfo {author}
  {\bibfnamefont {Y.}~\bibnamefont {Singh}}, \bibinfo {author} {\bibfnamefont
  {P.}~\bibnamefont {Gegenwart}},\ and\ \bibinfo {author} {\bibfnamefont
  {R.}~\bibnamefont {Valentí}},\ }\href
  {https://doi.org/10.1088/1361-648X/aa8cf5} {\bibfield  {journal} {\bibinfo
  {journal} {Journal of Physics: Condensed Matter}\ }\textbf {\bibinfo {volume}
  {29}},\ \bibinfo {pages} {493002} (\bibinfo {year} {2017})}\BibitemShut
  {NoStop}%
\bibitem [{\citenamefont {Motome}\ \emph {et~al.}(2020)\citenamefont {Motome},
  \citenamefont {Sano}, \citenamefont {Jang}, \citenamefont {Sugita},\ and\
  \citenamefont {Kato}}]{Motome2020IOP}%
  \BibitemOpen
  \bibfield  {author} {\bibinfo {author} {\bibfnamefont {Y.}~\bibnamefont
  {Motome}}, \bibinfo {author} {\bibfnamefont {R.}~\bibnamefont {Sano}},
  \bibinfo {author} {\bibfnamefont {S.}~\bibnamefont {Jang}}, \bibinfo {author}
  {\bibfnamefont {Y.}~\bibnamefont {Sugita}},\ and\ \bibinfo {author}
  {\bibfnamefont {Y.}~\bibnamefont {Kato}},\ }\href
  {https://doi.org/10.1088/1361-648X/ab8525} {\bibfield  {journal} {\bibinfo
  {journal} {Journal of Physics: Condensed Matter}\ }\textbf {\bibinfo {volume}
  {32}},\ \bibinfo {pages} {404001} (\bibinfo {year} {2020})}\BibitemShut
  {NoStop}%
\bibitem [{\citenamefont {Liu}\ \emph {et~al.}(2020)\citenamefont {Liu},
  \citenamefont {Chaloupka},\ and\ \citenamefont {Khaliullin}}]{Liu2020PRL}%
  \BibitemOpen
  \bibfield  {author} {\bibinfo {author} {\bibfnamefont {H.}~\bibnamefont
  {Liu}}, \bibinfo {author} {\bibfnamefont {J.~c.~v.}\ \bibnamefont
  {Chaloupka}},\ and\ \bibinfo {author} {\bibfnamefont {G.}~\bibnamefont
  {Khaliullin}},\ }\href {https://doi.org/10.1103/PhysRevLett.125.047201}
  {\bibfield  {journal} {\bibinfo  {journal} {Phys. Rev. Lett.}\ }\textbf
  {\bibinfo {volume} {125}},\ \bibinfo {pages} {047201} (\bibinfo {year}
  {2020})}\BibitemShut {NoStop}%
\bibitem [{\citenamefont {Winter}(2022)}]{Winter2022JPM}%
  \BibitemOpen
  \bibfield  {author} {\bibinfo {author} {\bibfnamefont {S.~M.}\ \bibnamefont
  {Winter}},\ }\href {https://doi.org/10.1088/2515-7639/ac94f8} {\bibfield
  {journal} {\bibinfo  {journal} {Journal of Physics: Materials}\ }\textbf
  {\bibinfo {volume} {5}},\ \bibinfo {pages} {045003} (\bibinfo {year}
  {2022})}\BibitemShut {NoStop}%
\bibitem [{\citenamefont {Liu}\ and\ \citenamefont {Kee}(2023)}]{XLiu2023PRB}%
  \BibitemOpen
  \bibfield  {author} {\bibinfo {author} {\bibfnamefont {X.}~\bibnamefont
  {Liu}}\ and\ \bibinfo {author} {\bibfnamefont {H.-Y.}\ \bibnamefont {Kee}},\
  }\href {https://doi.org/10.1103/PhysRevB.107.054420} {\bibfield  {journal}
  {\bibinfo  {journal} {Phys. Rev. B}\ }\textbf {\bibinfo {volume} {107}},\
  \bibinfo {pages} {054420} (\bibinfo {year} {2023})}\BibitemShut {NoStop}%
\bibitem [{\citenamefont {Rousochatzakis}\ \emph {et~al.}(2024)\citenamefont
  {Rousochatzakis}, \citenamefont {Perkins}, \citenamefont {Luo},\ and\
  \citenamefont {Kee}}]{Rouso2024RoPP}%
  \BibitemOpen
  \bibfield  {author} {\bibinfo {author} {\bibfnamefont {I.}~\bibnamefont
  {Rousochatzakis}}, \bibinfo {author} {\bibfnamefont {N.}~\bibnamefont
  {Perkins}}, \bibinfo {author} {\bibfnamefont {Q.}~\bibnamefont {Luo}},\ and\
  \bibinfo {author} {\bibfnamefont {H.-Y.}\ \bibnamefont {Kee}},\ }\href
  {https://doi.org/10.1088/1361-6633/ad208d} {\bibfield  {journal} {\bibinfo
  {journal} {Reports on Progress in Physics}\ }\textbf {\bibinfo {volume}
  {87}},\ \bibinfo {pages} {026502} (\bibinfo {year} {2024})}\BibitemShut
  {NoStop}%
\bibitem [{\citenamefont {Chaloupka}\ and\ \citenamefont
  {Khaliullin}(2016)}]{Chaloupka2016PRB}%
  \BibitemOpen
  \bibfield  {author} {\bibinfo {author} {\bibfnamefont {J.~c.~v.}\
  \bibnamefont {Chaloupka}}\ and\ \bibinfo {author} {\bibfnamefont
  {G.}~\bibnamefont {Khaliullin}},\ }\href
  {https://doi.org/10.1103/PhysRevB.94.064435} {\bibfield  {journal} {\bibinfo
  {journal} {Phys. Rev. B}\ }\textbf {\bibinfo {volume} {94}},\ \bibinfo
  {pages} {064435} (\bibinfo {year} {2016})}\BibitemShut {NoStop}%
\bibitem [{\citenamefont {Ozaki}(2003)}]{Ozaki2003PRB}%
  \BibitemOpen
  \bibfield  {author} {\bibinfo {author} {\bibfnamefont {T.}~\bibnamefont
  {Ozaki}},\ }\href {https://doi.org/10.1103/PhysRevB.67.155108} {\bibfield
  {journal} {\bibinfo  {journal} {Phys. Rev. B}\ }\textbf {\bibinfo {volume}
  {67}},\ \bibinfo {pages} {155108} (\bibinfo {year} {2003})}\BibitemShut
  {NoStop}%
\bibitem [{\citenamefont {Ozaki}\ and\ \citenamefont
  {Kino}(2004)}]{Ozaki2004PRB}%
  \BibitemOpen
  \bibfield  {author} {\bibinfo {author} {\bibfnamefont {T.}~\bibnamefont
  {Ozaki}}\ and\ \bibinfo {author} {\bibfnamefont {H.}~\bibnamefont {Kino}},\
  }\href {https://doi.org/10.1103/PhysRevB.69.195113} {\bibfield  {journal}
  {\bibinfo  {journal} {Phys. Rev. B}\ }\textbf {\bibinfo {volume} {69}},\
  \bibinfo {pages} {195113} (\bibinfo {year} {2004})}\BibitemShut {NoStop}%
\bibitem [{\citenamefont {Ozaki}\ and\ \citenamefont
  {Kino}(2005)}]{Ozaki2005July}%
  \BibitemOpen
  \bibfield  {author} {\bibinfo {author} {\bibfnamefont {T.}~\bibnamefont
  {Ozaki}}\ and\ \bibinfo {author} {\bibfnamefont {H.}~\bibnamefont {Kino}},\
  }\href {https://doi.org/10.1103/PhysRevB.72.045121} {\bibfield  {journal}
  {\bibinfo  {journal} {Phys. Rev. B}\ }\textbf {\bibinfo {volume} {72}},\
  \bibinfo {pages} {045121} (\bibinfo {year} {2005})}\BibitemShut {NoStop}%
\bibitem [{\citenamefont {Lejaeghere}\ and\ \citenamefont
  {et~al}(2016)}]{Lejaeghere2016Science}%
  \BibitemOpen
  \bibfield  {author} {\bibinfo {author} {\bibfnamefont {K.}~\bibnamefont
  {Lejaeghere}}\ and\ \bibinfo {author} {\bibnamefont {et~al}},\ }\href
  {https://doi.org/10.1126/science.aad3000} {\bibfield  {journal} {\bibinfo
  {journal} {Science}\ }\textbf {\bibinfo {volume} {351}},\ \bibinfo {pages}
  {aad3000} (\bibinfo {year} {2016})}\BibitemShut {NoStop}%
\bibitem [{\citenamefont {Weinberg}\ and\ \citenamefont
  {Bukov}(2019)}]{Weinberg2019SPP}%
  \BibitemOpen
  \bibfield  {author} {\bibinfo {author} {\bibfnamefont {P.}~\bibnamefont
  {Weinberg}}\ and\ \bibinfo {author} {\bibfnamefont {M.}~\bibnamefont
  {Bukov}},\ }\href {https://doi.org/10.21468/SciPostPhys.7.2.020} {\bibfield
  {journal} {\bibinfo  {journal} {SciPost Phys.}\ }\textbf {\bibinfo {volume}
  {7}},\ \bibinfo {pages} {020} (\bibinfo {year} {2019})}\BibitemShut {NoStop}%
\bibitem [{\citenamefont {Woodland}\ \emph
  {et~al.}(2023{\natexlab{b}})\citenamefont {Woodland}, \citenamefont
  {Macdougal}, \citenamefont {Cabrera}, \citenamefont {Thompson}, \citenamefont
  {Prabhakaran}, \citenamefont {Bewley},\ and\ \citenamefont
  {Coldea}}]{Woodland2023PRB1}%
  \BibitemOpen
  \bibfield  {author} {\bibinfo {author} {\bibfnamefont {L.}~\bibnamefont
  {Woodland}}, \bibinfo {author} {\bibfnamefont {D.}~\bibnamefont {Macdougal}},
  \bibinfo {author} {\bibfnamefont {I.~M.}\ \bibnamefont {Cabrera}}, \bibinfo
  {author} {\bibfnamefont {J.~D.}\ \bibnamefont {Thompson}}, \bibinfo {author}
  {\bibfnamefont {D.}~\bibnamefont {Prabhakaran}}, \bibinfo {author}
  {\bibfnamefont {R.~I.}\ \bibnamefont {Bewley}},\ and\ \bibinfo {author}
  {\bibfnamefont {R.}~\bibnamefont {Coldea}},\ }\href
  {https://doi.org/10.1103/PhysRevB.108.184416} {\bibfield  {journal} {\bibinfo
   {journal} {Phys. Rev. B}\ }\textbf {\bibinfo {volume} {108}},\ \bibinfo
  {pages} {184416} (\bibinfo {year} {2023}{\natexlab{b}})}\BibitemShut
  {NoStop}%
\end{thebibliography}%

\end{document}